\documentclass[showpacs,showkeys,preprintnumbers,nofootinbib,aps,prd,reprint,eqsecnum]{revtex4-1} 
\usepackage{graphicx}
\usepackage{amsmath}
\usepackage{array,booktabs,multirow}
\usepackage{verbatim}
\usepackage{amsmath,amsfonts,amssymb}
\usepackage{slashed}
\usepackage{hyperref}
\usepackage{natbib}
\usepackage[a]{esvect}

\usepackage[usenames,dvipsnames]{xcolor}
\usepackage{hyperref}
\hypersetup{
    colorlinks=true,    		
    linkcolor=BrickRed,        	
    citecolor=OliveGreen,     	
    filecolor=Fuchsia,     		
    urlcolor=Blue           		
}

\allowdisplaybreaks

\newcommand{\E}{\mathrm{e}}
\newcommand{\I}{{\rm i}}

\newcommand{\ab}{{ab}}

\newcommand{\sldc}{{SL(2,\mathbb{C})}}

\newcommand{\bra}[1]{\langle #1 |}
\newcommand{\ket}[1]{| #1 \rangle}

\newcommand{\bFIG}{\begin{figure}\begin{center}\includegraphics}
\newcommand{\eFIG}{\end{center}\end{figure}}
\newcommand{\be}{\begin{equation}}
\newcommand{\ee}{\end{equation}}

\begin{document}
\title{Radiative corrections to the Lorentzian EPRL-FK spinfoam graviton}
\author{Aldo Riello}
\email{aldo.riello@cpt.univ-mrs.fr}
\affiliation{Aix Marseille Universit\'e, CNRS, CPT, UMR 7332, 13288 Marseille, France  and Universit\'e de Toulon, CNRS, CPT, UMR 7332, 83957 La Garde, France}
\begin{abstract}
I study the corrections engendered by the insertion of a ``melon'' graph in the bulk of the first-order spinfoam used for the graviton propagator. I find that these corrections are highly non-trivial and, in particular, that they concern those terms which disappear in the Bojowald-Bianchi-Magliaro-Perini limit of vanishing Barbero-Immirzi parameter at fixed area ($\gamma\rightarrow0,\gamma j\sim cst$). This fact is the first realization of the often cited idea that the spinfoam amplitude receives higher order corrections under the refinement of the underlying two-complex.
\end{abstract}

\maketitle
\onecolumngrid

\section{Introduction}

In the background-independent quantum-gravity community, the interest in non perturbative effects is growing fast. On the one side,  boosted by important results in tensor models \cite{Gurau2011,Bonzom2011,Gurau2012,Kaminski2013}, a lot of progress has been made in the understanding of tensorial-group-field-theory (TGFT ) renormalization \cite{BenGeloun2012,BenGeloun2012a,Carrozza2012,Carrozza2013}, in particular in the context of 3d gravity; on the other side, the structure of a divergent graph has been studied for the first time within a four dimensional spinfoam model implementing the simplicity constraints. To be more specific, in \cite{Riello2013} it was found  that - under a certain number of hypothesis - the self-energy graph (or ``melon'' graph, to borrow a  TGFT expression) of the Lorentzian EPRL-FK model \cite{Engle2008} is logarithmically divergent in the spin cut-off and has a particular tensorial structure which corrects the usual trivial gluing.\footnote{More precisely, the statement holds for the geometrically non-degenerate sector, and ``Euclidean'' internal geometry. See the reference for further details.}  The aim of this work is to investigate some of the possible physical consequences of that calculation, by looking for corrections in the spinfoam graviton propagator \cite{Bianchi2006,Bianchi2009,Bianchi2011}. The work is organized as follows: in  \autoref{section2} I review the spinfoam graviton-propagator calculation in enough detail to make the ensuing discussion easy to follow. In  \autoref{section3}, I present the object of my studies, which will be carried out in \autoref{section4}. I critically discuss the results in \autoref{section5}.


\section{Spinfoam graviton propagator \label{section2}}

\subsection{The general-boundary picture}
The task of recovering $N$-point functions from a background independent theory is non trivial. To capture the basic difficulty it is enough to observe that in a formal expression  like
\be
\langle \phi(x_1)\cdots\phi(x_n)\rangle = \int D\phi\, \phi(x_1)\cdots\phi(x_n) \exp\I S[\phi]
\label{n_point_naive}
\ee
with both the measure and the action on the rhs taken to be diffeomorphism invariant, the lhs cannot be anything but a constant in the $\{x_i\}$. Some years ago, in a series of articles \cite{Rovelli2006,Modesto2005,Bianchi2006} it was shown how to tackle the problem via the general-boundary formalism \cite{Oeckl2003,Oeckl2008,Oeckl2012}. In a nutshell, the idea is that an expression like \autoref{n_point_naive} can be made sense of in terms of quantities where the path integral is performed in a (possibly) confined spacetime region $M$, while keeping the value of all the dynamical fields on the boundary $\partial M$ fixed:
\be
\langle \phi(x_1)\cdots\phi(x_n)\rangle_{M,\varphi} = \int_{\phi|_{\partial M}=\varphi} D\phi\, \phi(x_1)\cdots\phi(x_n) \exp\I S_M[\phi].
\label{n_point}
\ee
This prescription, which reminds of the original Feynman's path integral for non-relativistic quantum mechanics, is decisive since among the fields which are fixed on $\partial M$ there is also the gravitational field, with respect to which is now meaningful to talk about positions. Ultimately, the previous expression is diffeomorphisms \emph{invariant} only in the bulk, and the boundary can be thought as ``frozen'' by some measurement. This still allows the $N$-point function to be diffeomorphism \emph{covariant}. The general-boundary formalism is a generalization (and \textit{a priori} not a modification) of the usual path integral quantization \cite{Oeckl2012}: indeed, the original quantization prescription is recovered whence $M$ is taken to be the four dimensional space between to equal time hypersurfaces in Minkowski spacetime.


\subsection{Spinfoam realization}

As anticipated, these ideas can be adapted to the spinfoam quantization program. The main obstacle arises from the need of letting the continuous (field) picture talk to the discrete (spinfoam) one. Two, naturally related, form of discreteness enter the game: the discreteness of the boundary state, in the form of an $SU(2)$ spin network, and that of the bulk, in the form of an EPRL-FK spinfoam with given boundary. The choice of a particular (superposition of) spin networks as a boundary state is dictated by the boundary geometry and by the variety of scales one wants to describe:\footnote{Remark that the spin-network superpositions here considered involve only different colorings of the \textit{same} graph. Considering more general situations would bring into the problem new conceptual difficulties.} if only large scale modes of the boundary geometry and fields are of interest, then relatively small spin network will be sufficient, and their colorings (and superposition coefficients) will be such that to reproduce the boundary intrinsic and extrinsic geometry \cite{Rovelli2006}. The expansion in the number of spinfoam (bulk) vertices follows a similar logic and can be seen either as a refinement of the bulk discretization lattice capturing more and more degrees of freedom, or as a development in the perturbative GFT expansion. I briefly comment about the two prescriptions in the last section.\\

In \cite{Bianchi2006} it is discussed how this approach yields the expression for the spinfoam lowest order graviton two point function:
\be
{G}_{\bf q}^{abcd}({\bf x},{\bf y})=\frac{\sum_s W[s] \hat h^{ab}({\bf x})  \hat h^{cd}({\bf y}) \Psi_{\bf q}[s]}{\sum_s W[s] \Psi_{\bf q}[s]}.
\label{sf_gp}
\ee
Here, $\bf q$ represents the classical intrinsic and extrinsic boundary geometry, and $\Psi_{\bf q}$ is the quantum minimal-dispersion coherent state describing it. $\Psi_{\bf q}[s]:=\bra{s}\Psi_{\bf q}\rangle$ is its projection on the spin-network basis element $\ket{s}=\ket{\Gamma,j_l,i_n}$, where $\Gamma$ is an abstract spin-network graph, and $\{j_l\}$ ($\{i_n\}$)  are spins (intertwiners) labelling its links (nodes). $\hat h^{ab}({\bf x})$ is the spinfoam operator corresponding to the linearized gravitational field operator; it acts at the spin-network node corresponding to the point ${\bf x}$ on the 3d metric manifold $(\partial M,{\bf q})$. Finally, $W[s]$ is the (dynamical) amplitude associated to the spin network $s$. It can be calculated order by order in a vertex expansion. In this paper I focus in a one- and in a three-vertex spinfoam expansion.


\subsection{Lorentzian EPRL-FK boundary state construction}

\begin{figure}[h!]
\begin{center}
\includegraphics[width=.25\textwidth]{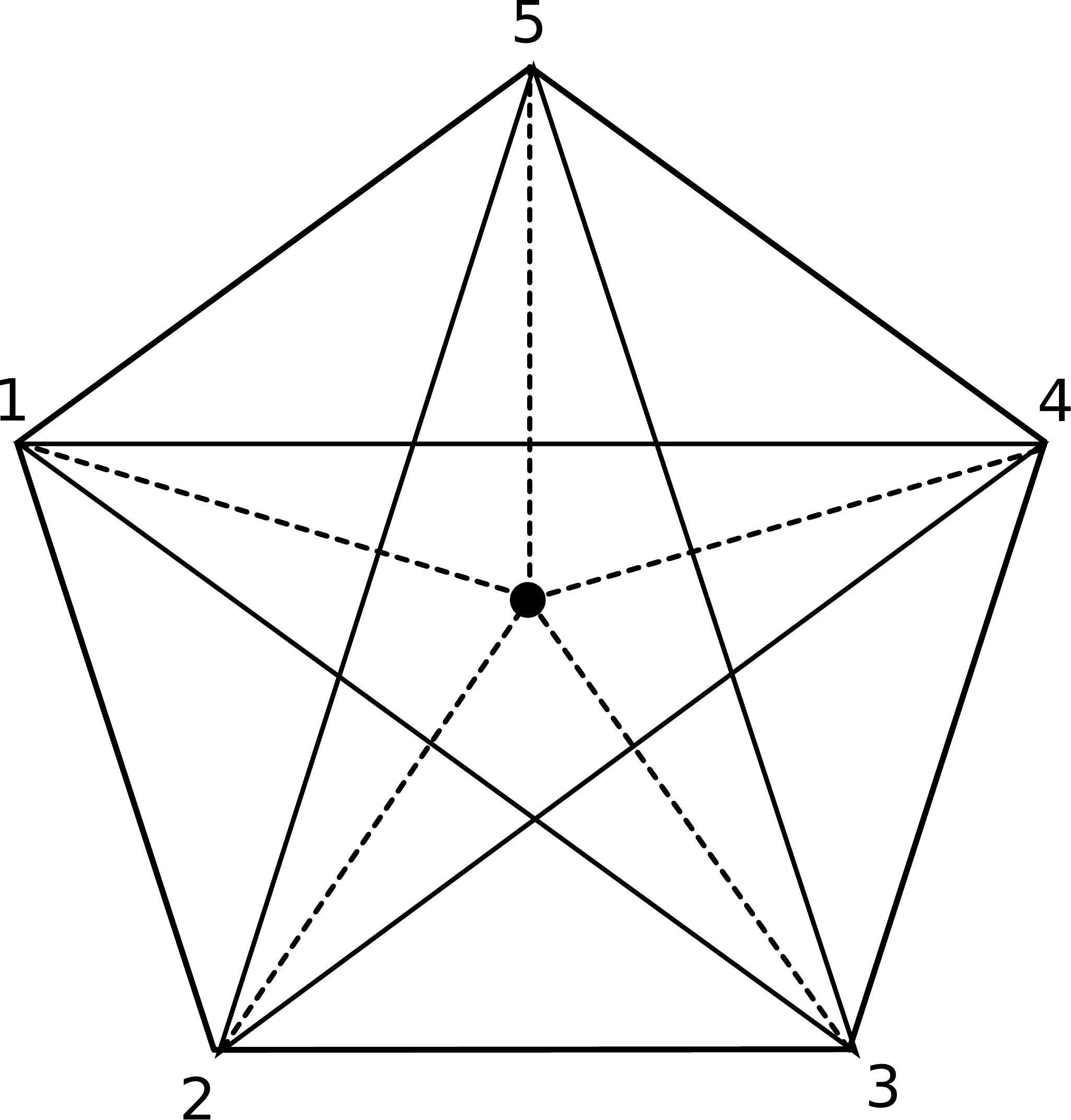}
\caption{The pentagon spinfoam $\mathcal C_5$.}
\label{penta}
\end{center}
\end{figure}

In the Lorentzian EPRL-FK model, the first non trivial order of \autoref{sf_gp} is given by a one-vertex spinfoam \cite{Bianchi2006}. Name this spinfoam the ``pentagon spinfoam'' $\mathcal C_5$, and its boundary the ``pentagon spin network'' $\partial \mathcal C_5=\Gamma_5$ (see \autoref{penta}).
The numbers $a\in\{1,\dots,5\}$ label the nodes of the spin network, solid lines represent its links (labelled by $\{ab,a<b\}$), while dashed lines represent the spinfoam edges (labelled by $\{a\}$), and the black dot the only spinfoam vertex.\\

The topology of the pentagon graph is that of a four-ball, with a three-sphere as a boundary. Since we are interested in the graviton propagator on flat spacetime, the four-ball should be thought of as a portion of four-dimensional Minkowski space. Also, to fit the usual Lorentzian EPRL-FK boundary states, the tetrahedra dual to the nodes must be taken space-like, i.e. with time-like normals.\\

In \cite{Bianchi2006,Bianchi2009,Bianchi2011} the construction of $\Psi_{\bf q}$ is carried out with great care. I briefly review the main results in order to fix the notation. At each node $a$ (for definiteness take $a=5$) consider a Livine-Speziale coherent intertwiner $\Phi_{a=5}$ \cite{Livine2007}:
\be
\Phi_{a=5}^{m_{1},\dots,m_{4}}(j_{15},\vec n_{15};\dots;j_{45},\vec n_{45}):=\int_{SU(2)} dh \, \prod_{b=1}^4\bra{j_{b5}, m_{b}}h\ket{j_{b5},\vec n_{b5}}.
\label{LS}
\ee
It is labelled by fur spins and four unit three-vectors satisfying a non-degenerate closure condition $\sum_{a=1}^4 j_a\vec n_a =\vec 0$. Taking all the normals to the tetrahedra outward pointing, there will be both future and past pointing tetrahedra. Supposing the tetrahedron $a=5$ is always future pointing, define
\begin{align}
\Upsilon_{a}:=\left\{
\begin{array}{ll}
\Phi_{a} & \text{if tetrahedron $a$ is future pointing}\\
\E^{-\I\sum_{b> a}\Pi_{ab}j_{ab}}\Phi_{a} & \text{if tetrahedron $a$ is past pointing}
\label{phases}
\end{array},
\right.
\end{align}
where\footnote{The ``thin'' and ``thick'' wedge terminology is standard. The wedge composed by two space-like tetrahedra is said to be ``thin'' if their normals are both future or past pointing, and ``thick'' otherwise.}
\begin{align}
\Pi_{ab}:=\left\{
\begin{array}{ll}
0 & \text{if wedge $(ab)$ is thick}\\
\pi & \text{if wedge $(ab)$ is thin}
\end{array}
\right. .
\end{align}

Naming $\upsilon^i_{m_1,\dots,m_4}$ the standard recoupling basis for intertwiners, define the coefficients
\be
\Upsilon_a^i(\{\vec n_{ab}\}):=\upsilon^i_{m_1,\dots,m_4}\Upsilon_a^{m_1,\dots,m_4}.
\ee
Hence, define the coherent Lorentzian spin network
\be
\ket{\Gamma_5,j_{ab},\Upsilon_a(\{\vec n_{ab}\})}:=\sum_{i_1,\dots,i_5}\left(\prod_{a=1}^5\Upsilon_a^{i_a}\right)\ket{\Gamma_5,j_{ab},i_a}.
\ee
This state is peaked on a given intrinsic geometry described by its labels, which have to be carefully chosen in such a way to guarantee its``geometricity". For the state to be peaked also on a given extrinsic geometry, one has to take a superposition of such coherent states\footnote{This is why these states are sometimes called \emph{semi}-coherent.} \cite{Rovelli2006,Bianchi2006}. Schematically:
\be
\ket{\Psi_o}=\sum_{j_{ab}}\psi_{j^o}(j)\ket{j,\Upsilon_a(\vec n)},
\ee
with coefficients $\psi_{j^o}(j)$ given by a Gaussian distribution times a complex phase:\footnote{This construction parallels that of standard quantum-mechanical coherent one-particle states, with wave function:
$$ \psi(x)\propto\exp\left( -\frac{(x-x_o)^2}{2\sigma_x^2} +\I {p^o x} \right).$$}
\begin{align}
&\psi_{j^o}(j)=\exp\left(-\I\sum_{ab}\gamma\phi_o^{ab}(j_{ab}-j_{ab}^o)
-\sum_{ab,cd}\gamma \alpha^{(ab),(cd)}\frac{j_{ab}-j_{ab}^o}{\sqrt{j_{ab}^o}}\frac{j_{cd}-j_{cd}^o}{\sqrt{j_{cd}^o}}\right) .
\label{coeff}
\end{align}
Here, $\phi^{ab}_o=\phi^{ab}_o(\{j^o\})$ is the simplicial extrinsic curvature, i.e. the dihedral angle, associated to the triangle $(ab)$ shared by the tetrahedra $a$ and $b$ within the four simplex defined by the triangle areas $j_{ab}^o$; the $10\times10$ matrix $\alpha$ is supposed to be complex with positive-defined real part. $\gamma$, the Barbero-Immirizi parameter, has been introduced for later convenience.


\subsection{The Lorentzian EPRL-FK graviton propagator}

The only missing piece of the spinfoam graviton-propagator construction is  the metric operator. For simplicity, and following \cite{Bianchi2006}, I work with the (densitized-)inverse-metric operator, which can be easily written in terms of the LQG flux operators through the triangles shared by the boundary tetrahedra. The flux operator at one boundary point, acts at one specific node $n$ along the link between the nodes $n$ and $a$. For each of the three tangential directions $i$, write this operator as $(E^a_n)^i$; therefore the inverse-metric operator reads:
\be
g^{ab}_n=\delta_{ij}(E^a_n)^i(E^b_n)^j.
\ee
When acting on a $SU(2)$ state $\ket{j_{cd},\vec n_{cd}}$ associated to a coherent Livine-Speziale intertwiner (\autoref{LS}), the flux operators acts as
\be
(E^a_n)^i \ket{j_{cd},\vec n_{cd}} = 8\pi G \hbar\gamma (\delta^a_c\delta^n_d+\delta^n_c\delta^a_d) J^i\ket{j_{cd},\vec n_{cd}},
\ee
where $G$ is the Newton constant and $J^i$ is the $i$-th component of the angular momentum operator in representation $j_{cd}$. \\

It is then a matter of calculations to show \cite{Bianchi2011} that the connected two point function of the metric operator at this order of perturbation theory in the Lorentzian EPRL-FK model can be recast into the path-integral form:
\be
G^{abcd}_{nm}=\frac{\sum_j \psi_j \int D^4g D^{10} z \,q_n^{ab} q_m^{cd} \E^S} {\sum_j \psi_j \int D^4g D^{10} z \, \E^S}-
\frac{\sum_j \psi_j \int D^4g D^{10} z \,q_n^{ab}  \E^S} {\sum_j \psi_j \int D^4g D^{10} z \, \E^S}\frac{\sum_j \psi_j \int D^4g D^{10} z \, q_m^{cd} \E^S} {\sum_j \psi_j \int D^4g D^{10} z \, \E^S},
\label{sf_gp_expl}
\ee
where $\psi_j$ are the coefficients of \autoref{coeff}; $D^4 g=\prod_{a=1}^5 Dg \delta(g_5)$ is the Haar measure over the $SL(2,\mathbb{C})$ bulk holonomies supplemented by the gauge fixing $g_5=\mathbb{I}$; $D^{10}z=\prod_{ab}Dz_{ab}$ is a scale invariant (spin-dependent) measure over 10 $\mathbb{CP}^1$ auxiliary variables (of which I should always consider the section of unit norm) needed to write the EPRL-FK amplitude in this path-integral form (see \cite{Barrett2010} for details). Define $\xi_{ab}=\ket{\frac{1}{2},\vec n_{ab}}\in\mathbb{C}^2$, then the insertion $q_n^{ab}$ can be written using the $\mathbb C^2$ Hermitian scalar product $\langle\cdot,\cdot\rangle$:
\be
q_n^{ab}=\delta_{ij} (A^a_n)^i(A^b_n)^j,\quad 
(A^a_n)^i=\gamma j_{an}\frac{\langle\sigma^i g_a^\dagger z_{an},\xi_{an}\rangle}{\langle g_a^\dagger z_{an}\xi_{an}\rangle},
\ee
$\sigma^i$ being the Pauli matrices. Finally the action $S$ is nothing but the usual vertex action augmented by the thin/thick wedge phases of \autoref{phases} (see \cite{Barrett2010} for details):\footnote{In \autoref{action}, $\mathcal{J}$ is the $SU(2)$ anti-linear structure map.}
\begin{align}
S(g,z)&=\sum_{a<b} S_{ab}(g_a,g_b,z_{ab})\\
S_{ab}&=j_{ab}\log\frac{\langle  \mathcal{J}\xi_{ab},g_a^\dagger z_{ab}  \rangle^2\langle  g_b^\dagger z_{ba},\xi_{ba}\rangle^2}{\langle g_a^\dagger z_{ab},g_a^\dagger z_{ab} \rangle\langle g_b^\dagger z_{ab},g_b^\dagger z_{ab}  \rangle}
+\I\gamma j_{ab}\log\frac{\langle g_b^\dagger z_{ab},g_b^\dagger z_{ab} \rangle}{\langle g_a^\dagger z_{ab},g_a^\dagger z_{ab}  \rangle}
-\I\Pi_{ab} j_{ab}.\label{action}
\end{align}

Now, in order to extract some intelligible physics out of \autoref{sf_gp_expl}, one can simplify the formulas by taking the limit of large distances. Remark that this is exactly the physical regime of interest in order to make contact with the graviton propagator on a semiclassical background far away from the Planck scale. Formally, this limit is simply achieved by uniformly rescaling all the spins appearing in \autoref{sf_gp_expl} by a common factor $\lambda\rightarrow\infty$:
\be
j_{ab}\mapsto\lambda j_{ab},\quad j^o_{ab}\mapsto\lambda j^o_{ab}.
\ee

However, before proceeding with the analysis of \autoref{sf_gp_expl} in the large distance regime, it is convenient to manipulate it one last time. Following \cite{Bianchi2011} once more, observe that a total effective action can be introduced, which scales linearly in $\lambda$:
\be
S_{tot}=S+\log\psi_j \mapsto \lambda S_{tot}.
\ee
Furthermore, in the large-spin limit, the discreteness of the spins themselves becomes less and less relevant and the sum over the spins can be substituted with an integral (at least close to the region where the integrand is peaked):
\be
\sum_j \mapsto \int D^{10} j.
\ee
In this way, \autoref{sf_gp_expl} can be formally written as\footnote{The $\lambda^{-4}$ factor in front of $G$ comes from the scaling $q_n^{ab}\mapsto \lambda^2q_n^{ab}$.}
\begin{align}
&\lambda^{-4}G^{abcd}_{nm}
=\frac{\int Dx \,q_n^{ab} q_m^{cd} \E^{\lambda S_{tot}}} {\int Dx \, \E^{\lambda S_{tot}}}-
\frac{\int Dx \,q_n^{ab} \E^{\lambda S_{tot}}} {\int Dx \, \E^{\lambda S_{tot}}}\frac{\int Dx \, q_m^{cd} \E^{\lambda S_{tot}}} {\int Dx \, \E^{\lambda S_{tot}}},
\end{align}
where $x$ summarizes all the 24 variables $(j,g,z)$.\\

Applying the stationary point approximation to this expression, while $\lambda\rightarrow\infty$, it is not difficult to realize\footnote{The idea is first found in \cite{Livine2006}.} \cite{Bianchi2009} that at leading order in negative powers of $\lambda$:
\be
\lambda^{-4}G^{abcd}_{nm}=\left. \lambda^{-1} (H^{-1})^{ij} (q_n^{ab})'_i (q_m^{cd})'_j \right|_{x_o}+O(\lambda^{-2}),
\label{G=Hinv}
\ee
where both the Hessian of the total action $H_{ij}=\partial^2 S_{tot}/\partial x^i \partial x^j$ and the partial derivatives of the insertions $q'_i=\partial q/\partial x^i$ are understood to be evaluated at the stationary point $x_o$. In turn, $x_o$ is defined as the solution of $\Re S_{tot}|_{x_o}=\sup \Re S_{tot}=0$ and $\partial S_{tot}/\partial x^i|_{x_o}=0$.\\

Remark that at this order of approximation the details of the measure $Dx$ do not play any role. This fact will be relevant in the next section.\\

To work out the structure of the Hessian of the total action, it's useful to take advantage of a clever parametrization of variables around the stationary point $x_o$ and of the fact that the action $S$ is linear in the spins. However, before entering into such details, let me recall what the stationary point looks like. The condition $\Re S_{tot}|_{x_o}=\sup \Re S_{tot}=0$ implies that both $\Re S|_{x_o}$ and $\Re\log \psi_j|_{x_o}$ are null.\footnote{This is the case because of the non-positivity of $\Re S$ (by Cauchy-Schwarz inequality) and of the positivity of the real part of the matrix $\alpha$.} In particular, the latter condition implies
\be
j_{ab}=j_{ab}^o.
\ee 
Since in building the superposition of spin-network states which is $\Psi_o$, care was taken of choosing the vectors $\vec n_{ab}$ as functions of the spins in such a way they always\footnote{In a neighbourhood of $\{j_{ab}^o\}$, at least.} describe a geometrical four simplex, so do the $\vec n_{ab}^o=\vec n_{ab}(\{j^o\})$. As shown in \cite{Barrett2010}, the stationary point equations for the action $S$ select, then, those unique holonomies $g^o_a(\{j^o\})$ which parallel transport the vectors $\vec n_{ab}^o$ (and the spinors $\xi_{ab}^o$) at the centre of the four-simplex and onto one another (following the four-simplex combinatorics, of course), as well as those $\mathbb{CP}^1$ variables $z_{ab}^o(\{j^o\})$ which are essentially\footnote{Actually, up to a crucial $U(1)$ phase and an irrelevant normalization. See the references for the details.} equal to $\mathcal{J} g^o_a \xi^o_{ab}$ (with $a<b$). Also, it turns out that the action $S$ evaluated at the stationary point $x_o$ is (numerically) equal to the Regge action of the four-simplex:
\be
S|_{x_o}=\I S_{Regge}(\{j^o\})=\I \sum_{ab} \gamma\, j_{ab}^o\,\phi^{ab}_o(\{j^o\}).
\label{Regge}
\ee
Notice that the phase choice of \autoref{coeff} is the only one which would have provided a solution to the equation $\partial S/\partial j|_{x_o}=0$. This can be easily seen by using the linearity of $S$ in the spins.\\

Back to the particular parametrization needed to simplify the form of the Hessian, the uniqueness of the solution I just described for a given value of the spins $j_{ab}$ suggests to parametrize the variables $g$ and $z$ in a neighbourhood of the stationary point in the following way:
\be
x=(j_{ab},g_a,z_{ab})=(j_{ab},h_a g_a^o(j_{ab}), z_{ab}^o(j_{ab})+\delta z_{ab})\mapsto x=(j_{ab},\beta^a_i,\delta z_{ab}),
\label{clever}
\ee
where $\sldc\ni h_a\approx\mathbb{I}+\beta^a_i\sigma^i$, and $\beta^a_i\in\mathbb{C}$. In  words, for any value of the spins $j_{ab}$ close to the critical one, the data $x$ at a point $\beta^a_i=0$, $\delta z_{ab}=0$ are always taken to describe a geometrical four simplex. This ``following the geometricity'' makes the form of the Hessian quite simple. In particular it sets $\partial^2 S/\partial j\partial \beta = \partial^2 S/\partial j\partial (\delta z) = 0$. Indeed, the previous parametrization assures that one gets simply zero when varying with respect to the spins the first variations of the action with respect to the $\beta$ (i.e. the $g$) and $\delta z$ , since these are nothing else than the geometricity conditions expressed in terms of closures and parallel transports \cite{Han2011}. Hence:
\be
H = \left(
\begin{array}{c|cc}
\phantom{\Big(} Q_{10\times10} & 0_{10\times24} & 0_{20\times20} \\\hline
\phantom{\Big(} 0_{24\times20} & H^{\beta,\beta}_{24\times24} &  H^{\beta,\delta z}_{24\times20}\\
\phantom{\Big(} 0_{0\times20} & H^{\delta z,\beta}_{20\times24} & H^{\delta z,\delta z}_{20\times20}
\end{array}\right)=
\left(\begin{array}{c|cc}
  \phantom{\Big(}  Q & 0 & 0  \\ \hline
 \phantom{\Big(}   0  & \multicolumn{2}{c}{\multirow{2}{*}{$X_{44\times44}$}} \\
 \phantom{\Big(}   0 & & 
  \end{array}\right),
\label{HQX}
\ee
where
\be
 Q_{(ab)(cd)}=\left.\frac{\partial^2 S_{tot}}{\partial j_{ab}\partial j_{cd}}\right|_{x_o}=-\frac{\gamma\alpha_{(ab)(cd)}}{\sqrt{j_{ab}^o j_{cd}^o}}+\I\frac{\partial^2 S_{Regge}}{\partial j_{ab}\partial j_{cd}}.
\label{Q}
\ee
Remark that also the last equality of the previous equation stems crucially from the choice of parametrization.\\

This concludes the leading order calculation of the graviton propagator in the EPRL-FK model. Indeed, the quantities $(q_n^{ab})'$ and $H$ can be calculated,\footnote{At least in principle: the previous parametrization is highly non trivial for practical purposes, and turns this otherwise easy task in a difficult problem.} and the Hessian can be (at least ideally) inverted. At this point, observe that it would have been more satisfactory if $H$ contained \textit{only} the second derivatives of the (area) Regge action appearing in $Q$. Indeed, the result
\be
G^{abcd}_{nm}\sim \sum_{(ef)(gh)}Q_{(ef)(gh)}^{-1} \frac{\partial q_n^{ab}}{\partial j_{ef}}\frac{\partial q_m^{cd}}{\partial j_{gh}}
\label{sought}
\ee
would match with the two-point function computed in perturbative Regge calculus with a boundary state \cite{Bianchi2006,Bianchi2008}. In the next subsection I review how, and in which sense, one can recover this result.

\subsection{The Bojowald-Bianchi-Magliaro-Perini limit}
As argued in a series of papers \cite{Bojowald2001,Bianchi2009,Magliaro2011}, in order to recover the classical limit of spinfoams (and loop quantum cosmology, and probably LQG in general) one should consider the limit $\gamma\rightarrow0$ while keeping $\gamma j_{ab}$ constant and large. The meaning of this limit is to send to zero the quantum geometry effects by sending to zero the area gap between the area eignevalues, which in the large-spin limit is $\sim8\pi G\hbar\gamma/2$, while keeping the geometry, i.e. the area scale of the problem which in the same limit is $\sim8\pi G\gamma j$, macroscopic. Call this the Bojowald-Bianchi-Magliaro-Perini (BBMP) limit. In a quantum-geometrical context, this limit is somewhat similar to that of a rotating object made semiclassical by taking $\hbar\rightarrow0$ while keeping the total angular momentum $L=\hbar l$ finite and large.\\

In order to apply the BBMP limit to the spinfoam graviton problem, one must first investigate the dependence of the Hessian $H$ from both  the Barbero-Immirzi parameter and the spin scale $j$, defined by setting $j_{ab}=j \epsilon_{ab}$, with $\epsilon_{ab}\sim O(1)$. A careful analysis, and many calculations, show the following result \cite{Bianchi2011}:
\be
Q =  \frac{\gamma j}{j^2} Q_\epsilon
\quad \text{and}\quad
X = j (X_\epsilon + O(\gamma)),
\ee
where $Q_\epsilon,\,X_\epsilon$ depend only on $\epsilon$ (and neither on $\gamma$ nor $j$). Therefore, in the BBMP limit, the inverse of $H$ is:
\be
H^{-1} \stackrel{\text{BBMP}}{=}
\left(\begin{array}{c|c}
    \phantom{\Big(} j^2(\gamma j)^{-1} Q_\epsilon^{-1} & 0  \\ \hline
    0  &\phantom{\Big(} j^{-1} (X_\epsilon^{-1}+O(\gamma))
  \end{array}\right).
\ee
Consider now the derivatives of the isertions $(q_n^{ab})'$. A moment of reflection shows that they scale as (see \cite{Bianchi2011} for more details)
\be
\frac{\partial q}{\partial j} =  \gamma (\gamma j) {(q')}^\epsilon_j \sim O(\gamma (\gamma j)), 
\quad
\frac{\partial q}{\partial \beta} = (\gamma j)^2  {(q')}^\epsilon_\beta \sim O\left((\gamma j)^2\right),
\quad \text{and}\quad
\frac{\partial q}{\partial \delta z} =(\gamma j)^2 {(q')}^\epsilon_z  \sim O\left((\gamma j)^2\right).
\ee
Hence, schematically
\begin{align}
G &\stackrel{\text{BBMP}}{=} (\gamma j)^3 Q_\epsilon^{-1}{(q')}^\epsilon_j {(q')}^\epsilon_j + O\left(j^{-1}(\gamma j)^4 \right) + O\left(\gamma j^{-1}(\gamma j)^4 \right) \nonumber\\
& \stackrel{\text{BBMP}}{=} (\gamma j)^3 \left[ Q_\epsilon^{-1}{(q')}^\epsilon_j {(q')}^\epsilon_j  + O(\gamma)  \right].
\label{final}
\end{align}

So, the sought semiclassical result (\autoref{sought}) was shown to arise from the BBMP limit. Remark, that the correlations expressed in \autoref{final} scale with the inverse squared distance between the two nodes, i.e. as $\sim(\gamma j)^{-1}$. To realize this one has to recall the fact that each of the correlated objects (the $q$'s) scales by itself as $(\gamma j)^2$ (see \cite{Bianchi2009}).\\

A faster way to get directly to this formula, would have been doing the change of variables $\gamma j_\ab \mapsto k_\ab$, in order to take the limit $\gamma\rightarrow0$ directly at the level of \autoref{sf_gp_expl}. As observed in \cite{Magliaro2011}, a leading order saddle point appoximation of this limit automatically projects the EPRL-FK spinfoam calculations to a specific quantum Regge calculus, where the only variable are the $k_{ab}$. Then,  the limit $k_{ab}\rightarrow\infty$ would be simply the semiclassical limit of the quantum Regge calculus and would  directly lead to the previous result. Obviously, in this way, one losses the possibility of investigating the corrections in the Barbero-Immirzi parameter.\\

To conclude this section, I point out a recent series of papers by Muxin Han which pushes the ideas related to the BBMP limit to a higher level of precision and sophistication \cite{Han2013,Han2013a,Han2013b}, clarifying some aspects of the low-energy limit of the EPRL-FK model.


\section{Radiative corrections to the pentagon spinfoam\label{section3}}

\begin{figure}[h!]
\begin{center}
\includegraphics[width=.25\textwidth]{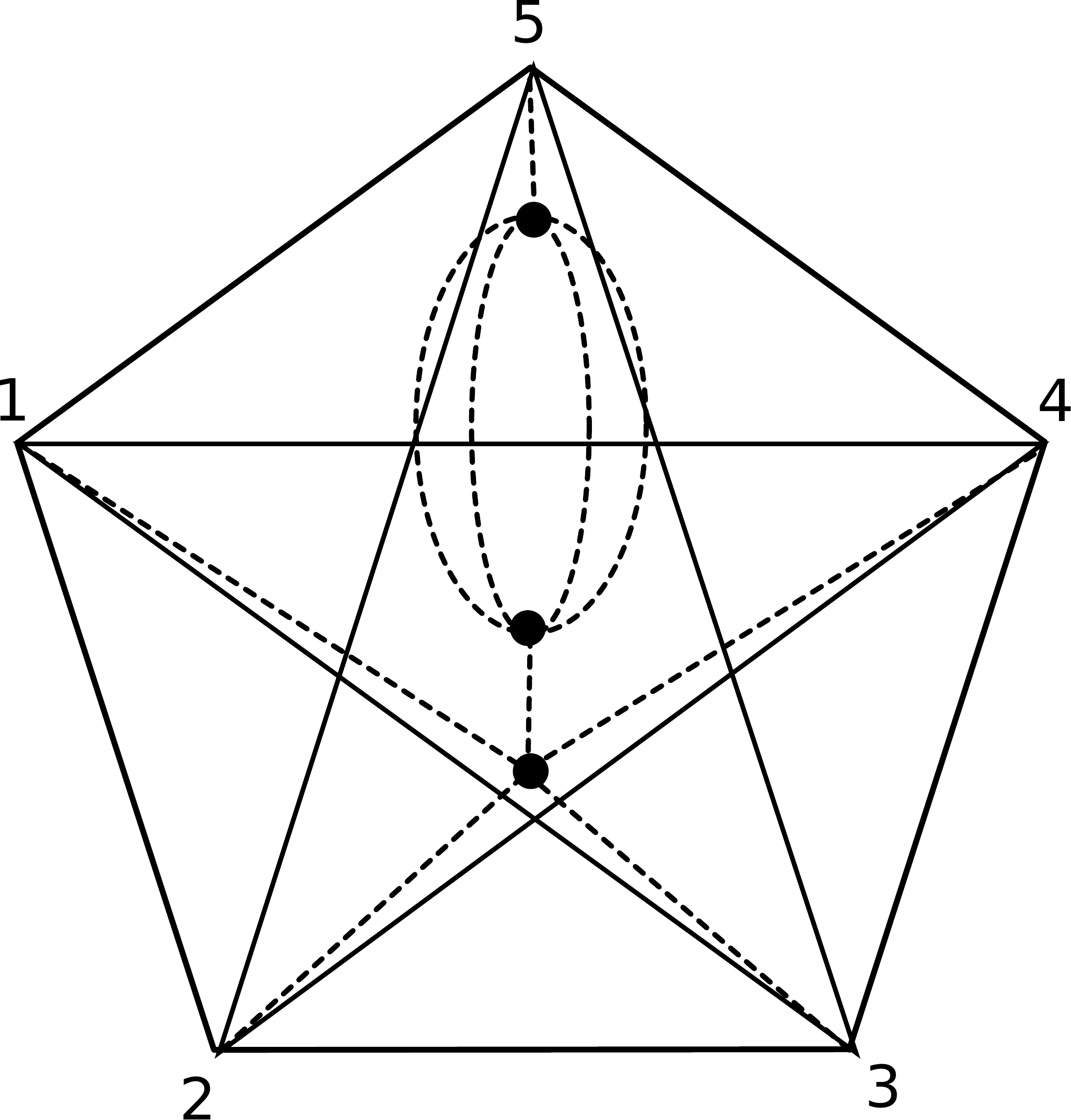}
\caption{The bubble-corrected pentagon spinfoam $\mathcal C_{\mathcal M}$.}
\label{pentamelon}
\end{center}
\end{figure}

After having reviewed the calculation of the first-order spinfoam graviton propagator, it's time to study the main subject of this work, i.e. the corrections to the previous calculation engendered by the presence of a ``melonic'' bubble on one of the edges of $\mathcal C_5$ (see \autoref{pentamelon}). Name this new spinfoam $\mathcal C_{\mathcal M}$.\\

\begin{figure}[h!]
\begin{center}
\includegraphics[width=.25\textwidth]{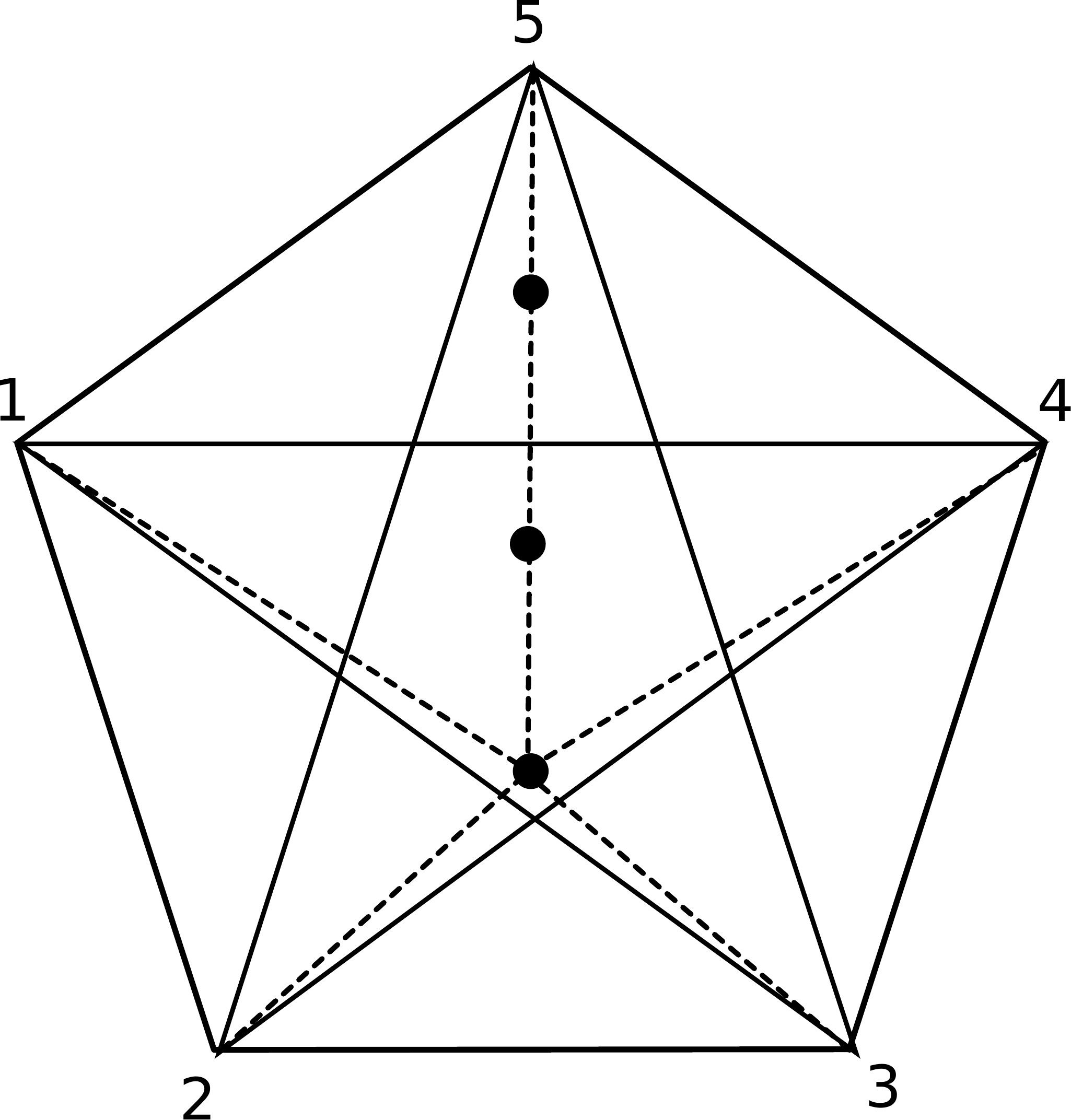}
\caption{The spinfoam $\mathcal C'_5$.}
\label{C3}
\end{center}
\end{figure}

There are two types of faces in $\mathcal C_{\mathcal M}$: external and internal. The external faces have spin fixed by the boundary state. The internal faces have by definition unconstrained spins and, in this case, all of them are two-edge-long faces confined within the melonic insertion. These unconstrained spins must be summed over. The (approximative) result of this procedure was worked out in \cite{Riello2013}, and is given by the following:
\be
W_{C_\mathcal{M}}\sim \log\left(\frac{\Lambda}{\ell}\right) W_{\mathcal C'_5} + \text{finite terms},
\label{reduction}
\ee
where $\mathcal C'_5$ is given in \autoref{C3}, $\Lambda$ is a cut-off on the spins\footnote{In \cite{Riello2013} it was claimed that $\Lambda$ can be qualitatively interpreted as the inverse cosmological constant, hence playing the role of a physical cut-off. This interpretation is not viable here, since we are dealing with propagation on a Minkowskian background, and not on a de Sitter one. To be more precise, the calculation is valid only at scales $j^o\ll\Lambda$, where the cosmological curvature can be neglected. This physical restriction is the same as the condition which is mathematically needed to make sense of the approximative result of \cite{Riello2013} used to deduce \autoref{reduction}. See the cited reference for more details.} (to be taken $\Lambda\gg \ell$), and $\ell$ is the scale of the boundary spins (morally $\ell\sim\langle j^o_{ab} \rangle$). The exact dependence of the coefficient in front of the logarithm and of $\ell$ from the boundary spins is not known.

It is now enough to go through the whole procedure described in the previous section, just substituting the amplitude $W_{\mathcal C_5}$ with $\left[\log\left({\Lambda}/{\ell}\right) W_{\mathcal C'_5}\right]$. In particular, the boundary state on $\partial C_\mathcal{M}=\Gamma_5$ will be the same as before.\\

\begin{figure}[h!]
\begin{center}
\includegraphics[width=.33\textwidth]{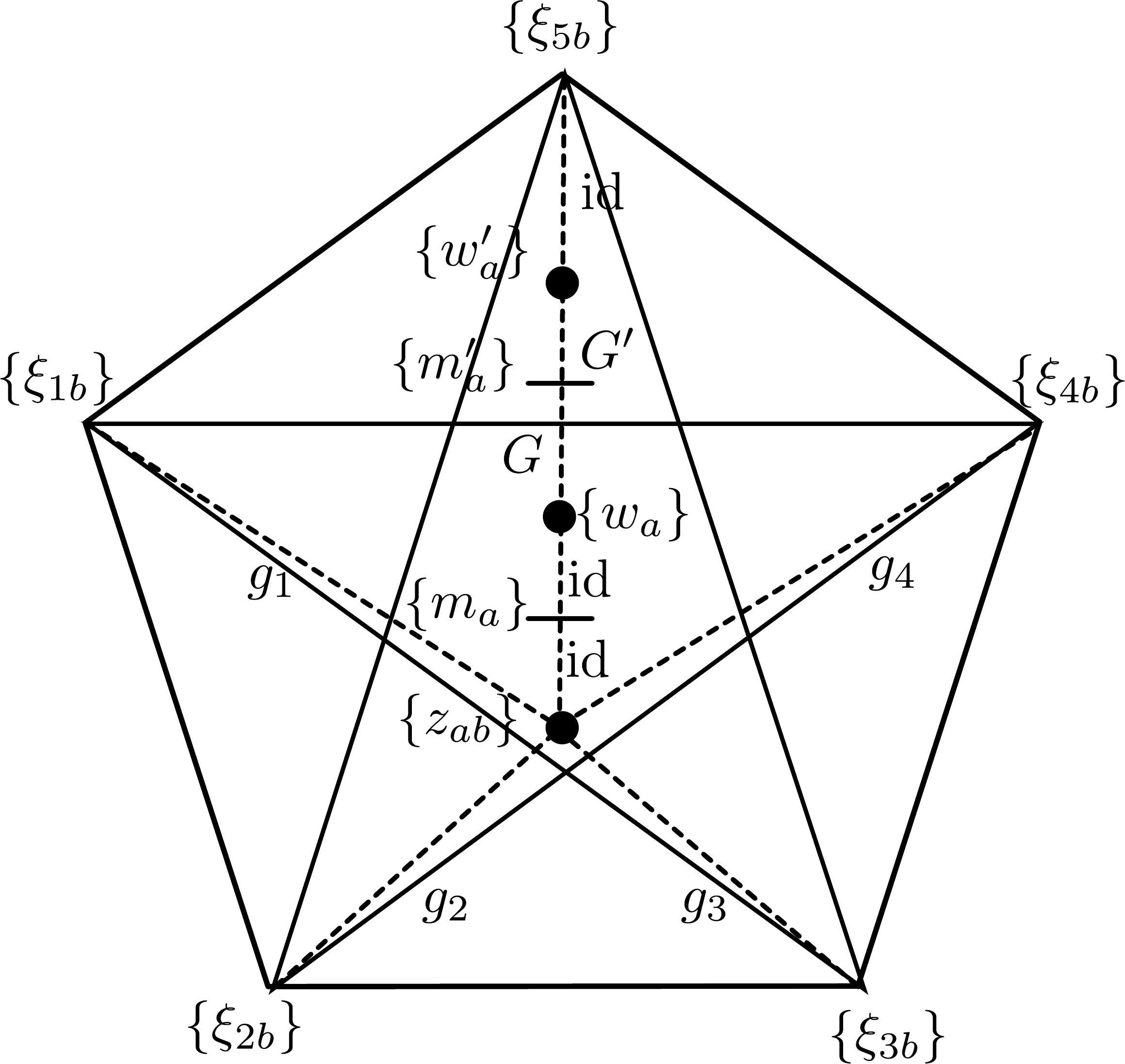}
\caption{The spinfoam $\mathcal C'_5$, where the variables living on it were highlighted.}
\label{C3_labels}
\end{center}
\end{figure}

The first important remark is that the (divergent) normalization $\log\left({\Lambda}/{\ell}\right)$ - in the same approximation as in the previous section - just modifies the ``path-integral'' measure $Dx$ by a $j$-dependent factor. As observed in the paragraph following \autoref{G=Hinv}, the details of this measure do not influence the leading order result I am going to investigate.\footnote{Though, they \textit{do} influence sub-leading orders.} Nevertheless, the presence of two more vertices in $W_{\mathcal C'_5}$ has some consequences even at this order. Specifically, more variables are needed to write the spinfoam amplitude and the action gets consequently modified:
\begin{align}
S'(g,z)&=\sum_{a<b} S'_{ab}(g_a,g_b,z_{ab})\\
a,b\neq 5, &\quad S'_{ab}=j_{ab}\log\frac{\langle  \mathcal{J}\xi_{ab},g_a^\dagger z_{ab}  \rangle^2\langle  g_b^\dagger z_{ba},\xi_{ba}\rangle^2}{\langle g_a^\dagger z_{ab},g_a^\dagger z_{ab} \rangle\langle g_b^\dagger z_{ab},g_b^\dagger z_{ab}  \rangle}
+\I\gamma j_{ab}\log\frac{\langle g_b^\dagger z_{ab},g_b^\dagger z_{ab} \rangle}{\langle g_a^\dagger z_{ab},g_a^\dagger z_{ab}  \rangle}
-\I\Pi_{ab} j_{ab}.\nonumber\\
b=5, &\quad S'_{a5}=j_{a5}\log\frac{\langle  \mathcal{J}\xi_{a5},g_a^\dagger z_{a5}  \rangle^2\langle   z_{5a},m_{a}\rangle^2}{\langle g_a^\dagger z_{a5},g_a^\dagger z_{a5} \rangle}
+\I\gamma j_{a5}\log\frac{1}{\langle g_a^\dagger z_{a5},g_a^\dagger z_{a5}  \rangle}\nonumber\\
&\qquad\qquad +j_{a5}\log\frac{\langle  m_{a}, w_{a}  \rangle^2\langle   G^\dagger w_{a},m'_{a}\rangle^2}{\langle G^\dagger w_{a},G^\dagger w_{a} \rangle}
+\I\gamma j_{a5}\log\langle G^\dagger w_{a},G^\dagger w_{a} \rangle\nonumber\\
&\qquad\qquad +j_{a5}\log\frac{\langle  m'_{a},(G')^\dagger w'_{a}  \rangle^2\langle   w'_{a},\xi_{5a}\rangle^2}{\langle (G')^\dagger w'_{a},(G')^\dagger w'_{a} \rangle}
+\I\gamma j_{a5}\log\frac{1}{\langle (G')^\dagger w'_{a},(G')^\dagger w'_{a}  \rangle}
-\I\Pi_{a5} j_{a5}.\label{Sa5}
\end{align}
where the gauge fixing was already taken into account (and ``concentrated'' along the row of edges stemming from node 5). Remark that it was necessary to introduce eight more (normalized) spinors $\{m_a,m'_a\}$, eight more $\mathbb{CP}^1$ variables $\{w_a,w'_a\}$ (taken in the unit-norm section), and two\footnote{Two other $\sldc$ elements have been tacitly gauge-fixed to the identity. See \autoref{C3_labels}. } more $\sldc$ variables $\{G,G'\}$ (see \autoref{C3_labels}). 

Finally, in order to calculate the correlation of the metric at the node 5 with the metric at some other node $a$, care should be taken of this adaptation of the insertion:
\be
q_5^{ab}=\delta_{ij} (A^a_5)^i(A^b_5)^j,\quad 
(A^a_5)^i=\gamma j_{a5}\frac{\langle\sigma^i w'_{a},\xi_{5a}\rangle}{\langle w'_{a},\xi_{5a}\rangle},
\ee
(while all other insertions stay unchanged).\\

Now, we have all the elements to carry out the saddle point analysis of the corrected graviton propagator, via the formula
\be
G^{abcd}_{nm}= \langle\langle q_n^{ab} q_m^{cd} \rangle\rangle- \langle\langle q_n^{ab}\rangle\rangle \langle\langle q_m^{cd} \rangle\rangle,\quad\text{where}\quad  \langle\langle Q\rangle\rangle = \frac{\sum_j \psi_j \int D^4g D^2 G D^{10} z D^8 w D^8 m \,Q\, \E^{S'}} {\sum_j \psi_j \int  D^4g D^2 G D^{10} z D^8 w D^8 m \, \E^{S'}}.
\ee

This will be the object of the next section.


\section{Saddle point analysis of the corrected graviton propagator\label{section4}}

The analysis of the stationary point of the new action is only slightly modified by the presence of the two new vertices. Indeed, as shown explicitly in \autoref{appendix1}, the group elements $G$ and $G'$ at the stationary point have to be in the $SU(2)$ subgroup of $\sldc$, making them almost irrelevant, since their effect can always be reabsorbed in other variables redefinition. This conclusion could be reached more directly from a result by Puchta \cite{Puchta2013} about the asymptotic (large-spin) form of the divergent part of the melon graph. A possible restatement of his result is that in the large-spin limit, one can ignore any insertion of additional vertices along a single edge. This is what happens to the two additional vertices of \autoref{C3_labels} with respect to \autoref{penta}. Therefore, the stationary point geometry is always that of the four simplex induced by the boundary state.\\

Then, what is more compelling to study is the way correlations between insertions behave, and whether or not differences arise with respect to the spinfoam $\mathcal{C}_5$. At a practical level, what one has to study is therefore the Hessian $H'$ of the action $S'$.\\

The first thing to notice is that $H'$ can be made block diagonal, therefore isolating the spin-spin correlations, by choosing a parametrization of the space around a stationary point which follows the ``geometricity'' conditions. This follows exactly the discussion of the previous section (in particular, see the paragraph of \autoref{clever}). For what concerns the other variables, name the ``old'' ones existing also in $\mathcal{C}_5$ (i.e. $\{g_a,z_{ab},\}$) $x'$, and the new ones (i.e. $\{w_a,w'_a,m_a,m'_a,G,G'\}$) $y$. So, the Hessian $H'$  can be schematically written as
\be
H'=
\left(\begin{array}{c|cc}
  \phantom{\Big(}  Q' &  0 & 0 \\\hline
  \phantom{\Big(}  0 &  X' & Z \\
  \phantom{\Big(}  0 &  Z^T & Y 
\end{array}\right),
\ee
where
\be
Q'=\left.\frac{\partial^2 S'_{tot}}{\partial j \partial j}\right|_{x_o},\quad X'=\left.\frac{\partial^2 S'_{tot}}{\partial x' \partial x'}\right|_{x_o},\quad Z=\left.\frac{\partial^2 S'_{tot}}{\partial x' \partial y}\right|_{x_o},\quad\text{and}\quad Y=\left.\frac{\partial^2 S'_{tot}}{\partial y \partial y}\right|_{x_o}.
\ee
A moment of reflection reveals that\footnote{After having carefully parametrized the space as discussed above.} $Q'=Q$ and $X'=X$ (see \autoref{HQX}). On the other hand, the matrix $Y$, can be worked out only by looking at the actions $S'_{a5}$ for the four faces $\{(a5)\}$ and is \emph{a priori} quite complicated. Of main interest, the matrix $Z$ has quite a simple form since it contains many null elements. Before analysing this matrix, notice that 
\be
\left.\frac{\partial q_{a}}{\partial x'}\right|_{x_o}\neq 0\;\text{and}\;\left.\frac{\partial q_5}{\partial x'}\right|_{x_o}=0,
\quad\text{while}\quad
\left.\frac{\partial q_a}{\partial x'}\right|_{x_o}=0 \;\text{and}\;\left.\frac{\partial q_5}{\partial y}\right|_{x_o}\neq0,
\ee
with $a\in\{1,\dots,4\}$.\\

Looking at $S'_{a5}$ in \autoref{Sa5}, it is immediate to see that the only non-null terms in $Z$ are
\be
Z_{z_{5a},m_a}=\left.\frac{\partial^2 S'_{tot}}{\partial z_{5a} \partial m_a}\right|_{x_o}.
\ee
Unluckily, these terms are enough to waste any simple attempt to reduce (even part of) the inverse of this Hessian ${(H')}^{-1}$ to $H^{-1}$. Indeed, it is easy to realize that the variables $m_a$  ($z_{a5}$) are - more or less directly - correlated to essentially all the other $x'$ (respectively $y$). In turn, this means that all the variables $\{x',y\}$ are cross-correlated among them. Therefore, abandoning for the rest of the paper the hope of any finer analysis, define
\be
\left(\begin{array}{cc}
 \multicolumn{2}{c}{\multirow{2}{*}{$Y'$}}\\
 &\\
\end{array}\right)=
\left(\begin{array}{cc}
  X'   \phantom{\Big(}& Z \\
 Z^T   \phantom{\Big(} & Y\\
\end{array}\right).
\ee

It is simple to see that the scale dependence of $Q'$ and $Y'$ from $j_o$ and $\gamma j_o$ (I have the BBMP limit in mind) is the same as that of $Q$ and $X$. This fact allows to immediately affirm that at leading order in the BBMP limit, the graviton propagator gets no modifications from the insertion of the melonic bubble within one of its edges. Once more, as discussed at the end of \autoref{section3}, this could have been directly deduced by making first the change of variables $\gamma j_{ab}\rightarrow k_{ab}$ followed by the two limits (taken in this order) $\gamma\rightarrow0$ and $k_{ab}\rightarrow\infty$.


\section{Summary and Discussion\label{section5}}

I have reviewed in some detail the calculation of the Lorentzian EPRL-FK spinfoam graviton propagator on the spinfoam $\mathcal{C}_5$ at leading order in an expansion over the external spin scale. I recalled that the expected classical result can be recovered in the the Bojowald-Bianchi-Magliaro-Perini semiclassical (BBMP) limit. I have then gone through the same calculation steps after inserting a ``melonic'' radiative correction to one of the spinfoam edges.\\

To deal with the new spinfoam, I first considerably simplified the problem by applying the result on the dominant order of the melon graph obtained in \cite{Riello2013}. In this way it was possible to effectively ``contract'' the bubble to two spinfoam vertices while forgetting its internal structure. The divergent factor this ``contraction'' comes with is irrelevant at the considered order of approximation. As a side remark, observe that this happens because of the normalization scheme used, i.e. ultimately because of the denominator of \autoref{sf_gp}. \\

Then, I noticed that the dominating process (which can be identified with the stationary point of the path integral form of the amplitude) is geometrically equivalent to the original one. Nonetheless, the correlation matrix of graviton fluctuations gets modified by the presence of the bubble. Moreover, being the correlation matrix the \emph{inverse} of the Hessian of the effective action of the spinfoam, this happens in a non-trivial way: \emph{a priori}, \emph{all} the cross correlations are modified, and not only those involving the node ``close'' to where the bubble was inserted (node 5 in my notation).\\

Finally, I showed that such non-trivial modifications of the spinfoam graviton propagator involve only those terms which disappear in the BBMP limit ($\gamma\rightarrow0,\;\gamma j^o\sim cnst.$), and henceforth do not modify the final leading-order result. Since the calculation presented here is set from the beginning within the lattice perspective \cite{Rovelli2011}, it appears to support the consistency of the lattice-refining scheme to the non-perturbative limit of spinfoam gravity. It would be interesting, for comparison, to study the same problem in the alternative scheme \cite{Freidel2005,Oriti:Book,Krajewski2010}, namely within an \textit{ab initio} group-field-theoretical (GFT) approach to the graviton propagator.\\

To conclude, I stress a couple of points about the consistency of the melonic-bubble radiative corrections with respect to previous results and known physics. First, the consistency is obtained only in the semiclassical BBMP limit, where spins are large and the Barbero-Immirzi parameter is reciprocally small, and couldn't be obtained otherwise. The reason for this is two-fold: on the one hand the melonic bubble is equivalent to the trivial gluing only in the case of large external spins (which here happen to be the same boundary spins which must be large in order to make contact with known low-energy physics); on the other hand, leading-order corrections to the expected result are present in the $1/j$ expansion (before taking the $\gamma\rightarrow0$ limit) in both the pentagon and melonic foams, and  there is no reason for such corrections to be same in the two cases.\\

At last, precisely the fact that the leading corrections stemming from the presence of the melonic bubble affect those terms which disappear in the $\gamma\rightarrow0$ limit, is preventing us to advance any physical interpretation of such radiative correction, for the simple reason that it affects terms on their own of difficult interpretation.


\acknowledgements
The author thanks Claudio Perini for a technical clarification about the original graviton-propagator calculation, and in particular Carlo Rovelli  for many fruitful discussions and comments on a previous version of this paper.


\appendix

\section{$G$ and $G'$ are in $SU(2)$\label{appendix1}}

This section can be seen as an independent confirmation of the main result of \cite{Puchta2013}, in the context of the melon graph.\\

In the following, it is shown that the stationary phase equations imply that $G,G'\in SU(2)$. I will do the calculation explicitly only for $G$, but remark that every step would go through in the case of $G'$ as well, just by sending $m_a\mapsto \xi_{5a},\;w_a\mapsto w'_a$ and obviously $G\mapsto G'$. Notice also that the closure equations that hold for both $\{\vec m_a\}$ and $\{\vec \xi_a\}$ (see here below for the precise definitions), hold for different reasons in the two cases: for the $\{\vec m_a\}$ they are one of the stationary phase equations, while for the $\{\vec \xi_a\}$ they hold by construction of the boundary state.\\

For simplicity of notation let me set $j_{a5}\mapsto j_a$ and let me take all the faces oriented in the same direction with respect to the edge $a=5$. Then, the closure equations for the $\{\vec m_a\}$ and $\{\vec m'_a\}$ read (see e.g. \cite{Han2011}):
\be
\sum_a j_a \vec m_a = \vec 0 =\sum_a j_a \vec m'_a ,
\ee
where $(\vec m)^i = \bra{m}\sigma^i\ket{m}$, $\sigma^i$ being the $i$-th Pauli matrix. Posing $\sigma^0=\mathbb I$, generalize this identity to one among four-vector (see e.g.  \cite{Riello2013} for more technical details and notations):
\be
\sum_a j_a  m_a^I = A t^I =\sum_a j_a (m'_a)^I ,
\ee
where I defined $m_a^I = \bra{m}\sigma^I\ket{m}$, $A=\sum_a j_a$, and $t^I=(1,\vec 0)$.\\

Among the stationary phase equations, there are also the following parallel-transport equations (see e.g. \cite{Han2011}):
\be
\left\{
\begin{array}{l}
G\mathcal{J}m'_a = \lambda_a^{-1} \mathcal{J} m_a \quad\text{iff}\quad m'_a = \lambda_a G^\dagger m_a \\
G m'_a = \lambda_a m_a\\
m_a  = \E^{\I\chi_a} w_a
\end{array}
\right.,
\ee
where $\lambda_a=||G m'_a|| \E^{\I \psi_a}$ and $\psi_a,\chi_a\in[0,2\pi]$.\\

Now, the goal is to show that $G\in SU(2)$ in order for it to properly parallel transport the closure relations. Hence, first use the fact that $(\mathcal{J}m_a)^I=\mathcal P \rhd m_a^I$ ($\mathcal P$ being the parity reversal operator) to calculate
\be
G\rhd At^I=
\left\{
\begin{array}{l}
 G\rhd\sum_a j_a (m'_a)^I =\sum_a j_a (Gm'_a)^I= \sum j_a |\lambda_a|^{2} m_a^I\\\\
 G\rhd\sum_a j_a (\mathcal{J} m'_a)^I =\sum_a j_a (G\mathcal{J} m'_a)^I= \sum j_a |\lambda_a|^{-2} (\mathcal{J} m_a)^I=\mathcal{P}\rhd \sum j_a |\lambda_a|^{-2} m_a^I\\
\end{array}
\right. .
\ee
Then, comparing the two last terms on the rhs of this equation, it is immediate to deduce that
\be
\sum_a j_a\left( |\lambda_a|^2+\frac{1}{|\lambda_a|^2}  \right)\vec m_a=\vec 0
\ee
However, since the tetrahedra defined by the four face vectors $\{j_a\vec m_a\}$ are non-degenerate\footnote{The non degeneracy conditions follows from the fact that the tetrahedra are defined up to rotations and reflections only by the spins $\{j_a\}$ which are the same defining the tetrahedra on the boundary of the spinfoam, which in turn are non-degenerate by hypothesis.} (i.e. any triple taken out of these vectors spans $\mathbb R^3$), it must be
\be
|\lambda_a|^2+\frac{1}{|\lambda_a|^2} \equiv C
\label{C}
\ee
$C$ being a constant independent from the index $a$, which is readily shown to be $C\geq2$.\\

I now claim that if $|\lambda_a|=1$, i.e. $C=2$, then $G\in SU(2)$. To show this, first observe that from the parallel-transport equations it follows that
\be
\left\{
\begin{array}{l}
\langle w_a, GG^\dagger w_a\rangle = |\lambda_a|^2=1\\
\langle \mathcal{J}w_a, GG^\dagger \mathcal{J}w_a\rangle = |\lambda_a|^{-2}=1\\
\langle \mathcal{J}w_a, GG^\dagger w_a\rangle \propto_\mathbb{C} \langle \mathcal{J}w_a, Gm'_a\rangle  \propto_\mathbb{C} \langle \mathcal{J}w_a, m_a\rangle \propto_\mathbb{C} \langle \mathcal{J}w_a, w_a\rangle \equiv 0
\end{array}
\right. .
\ee
Then, since $\{w_a,\mathcal{J}w_a\}$ is an orthonormal basis of $\mathbb{C}^2$ for any $a$, this shows $GG^\dagger=\mathbb I$, i.e. $G\in SU(2)$.\\

Therefore, I only need to show that $|\lambda_a|=1$. To get to this result, the idea is to compare the different orthonormal basis of $\mathbb C^2$ given by $\{w_a,\mathcal{J}w_a\}$ for different $a$. In particular, \autoref{C} implies
\be
|\lambda_a|^2=C^{(\pm)_a} \quad\text{where}\quad C^{\pm}= \frac{C\pm\sqrt{C^2-4}}{2},
\ee
hence, defining $\alpha,\beta\in\mathbb{C}$ via $w_1 = \alpha w_2 + \beta \mathcal{J}w_2$, so that $|\alpha|^2+|\beta^2|=1$, one gets e.g.:
\begin{align}
C^{(\pm)_1}&=|\lambda_1|^2=|| G^\dagger w_1||^2 = ||\alpha G^{\dagger}w_2 + \beta G^{\dagger} \mathcal{J}w_2||^2 \notag\\
&= |\alpha|^2 ||G^{\dagger}w_2||^2+|\beta|^2 ||G^{\dagger} \mathcal{J}w_2||^2 + 2\Re \left[  \alpha\bar\beta \langle G^{\dagger}w_2,G^{\dagger} \mathcal{J}w_2\rangle\right]\notag\\
&=|\alpha|^2 C^{(\pm)_2} + |\beta|^2 \frac{1}{C^{(\pm)_2}}.
\end{align}

Two cases are given: either $(\pm)_1=-(\pm)_2$ or $(\pm)_1=(\pm)_2$. In the first case, since $C^+ C^- =1$, the previous equation (together with the condition $C^{\pm}>0$) immediately implies $|\lambda_2|^2=C^{(\pm)_2}=1$. In the second case, the previous equation can be rewritten as
\be
(1-|\alpha|^2)\left[ \left(C^{(\pm)_2}\right)^2 -1\right] =0.
\ee
Remark that $|\alpha|=1$ implies $w_1 \propto_\mathbb{C} w_2$, which in turn implies $m_1 \propto_\mathbb{C} m_2$ and hence $\vec m_1 \propto \vec m_2$, which is in contradiction with the hypothesis of non-degeneracy of the tetrahedron. Therefore, also in this case one obtains $|\lambda_2|^2=C^{(\pm)_2}=1$. This concludes the demonstration.

\bibliographystyle{apsrev4-1}
\bibliography{biblio_melonic_graviton}

\end{document}